\documentclass[aps,twocolumn, superscriptaddress]{revtex4-1}
\usepackage[pdftex]{graphicx}% Include figure files
 \usepackage{amsmath}
\usepackage{amsfonts}
\usepackage[breaklinks=true, colorlinks=true, linkcolor= blue, urlcolor= blue, citecolor= blue]{hyperref}
\usepackage{tikz}
\usepackage{amssymb, mathrsfs}
\usepackage{braket}
\usepackage{subfigure}
\def\beq{\begin{equation}}
\def\eeq{\end{equation}}
\def\bsp{\begin{split}}
\def\esp{\end{split}}
\def\bea{\begin{eqnarray}}
\def\eea{\end{eqnarray}}
\def\ba{\begin{array}}
\def\ea{\end{array}}

\def\dg{\dagger}

\def\lb{\left(}
\def\rb{\right)}

\def\l.{\left.}
\def\r.{\right.}

\def\ra{\rangle}
\def\la{\langle}

\def\bo{\bold{k}}

\begin{document}

\title{Magnetic Order in Laser-Irradiated Kagom\'e Antiferromagnets}
\author{S. A. Owerre}
\affiliation{Perimeter Institute for Theoretical Physics, 31 Caroline St. N., Waterloo, Ontario N2L 2Y5, Canada.}

\begin{abstract}
Dispersionless   ``zero energy mode'' is one of the  hallmarks of frustrated kagom\'e antiferromagnets (KAFMs). It points to extensive classically  degenerate ground-states. The ``zero energy mode''  can be observed experimentally when lifted to a flat mode at finite energy by a strong intrinsic magnetic anisotropy. In this letter, we study the effects of irradiation of laser light on the KAFMs. We adopt the magnon picture without loss of generality.  It is shown that circularly or linearly polarized light lifts the ``zero energy mode'', stabilizes magnetic order, and induces energy gaps in the KAFMs. We find that the circularly  polarized light-induced  anisotropies have similar features as the intrinsic in-plane and out-of-plane Dzyaloshinskii-Moriya  interaction in KAFMs.  The former stabilizes long-range magnetic order and the latter induces spin canting out-of-plane with nonzero scalar spin chirality.  The Floquet thermal Hall effect shows that the synthetic magnetic excitation modes in the case of circularly  polarized light are topological, whereas those of linearly  polarized light are not.

\end{abstract}
\maketitle

Geometrically frustrated KAFMs have been an intensively  active field of research due to their exotic properties such as  the possibility of quantum spin liquids (QSLs) \cite{nor, zho, sav} --- states where spin frustration forbids   long-range magnetic  order down to the lowest temperatures.   Classically, the ideal Heisenberg KAFMs   have an  extensive  ground-state degeneracy \cite{chu, ch}, resulting in a  ``zero energy mode'' \cite{har1} --- a signature that no classical  ground-state configuration is favoured. However, order-by-disorder phenomenon \cite{chu, ch}  is believed  to lift the extensive classically  degenerate ground-states and selects a specific  magnetic  order, usually the ${\bf q =0}$ spin configuration in which three spins on the triangular plaquette of the kagom\'e lattice are oriented  at 120$^\circ$ apart.
Another possibility of lifting the extensive classically  degenerate ground-states  is by adding further neighbouring interactions \cite{har1}. Moreover, the geometry of the kagom\'e lattice lacks an inversion center  and by the Moriya's rules a  Dzyaloshinskii-Moriya  interaction (DMI) \cite{dm,dm2} is allowed. The out-of-plane DMI is capable of inducing   long-range magnetic  order in the frustrated  KAFMs \cite{men1}. Hence, the ``zero energy mode'' can be observed experimentally as a flat mode with finite energy \cite{men4a}.

In realistic quantum kagom\'e materials, however, further neighbouring interactions can  be perturbatively small and negligible, therefore the out-of-plane DMI  is usually  the dominant anisotropy in real materials.   Nevertheless,   magnetic  order in quantum spin-$1/2$ KAFMs appears beyond a certain quantum critical point (QCP) of the out-of-plane DMI ($D_z/J\sim 0.1$) \cite{men3}. Below the QCP it is believed that QSL persists.  For instance,  herbertsmithite ZnCu$_3$(OH)$_6$Cl$_2$ has dominant out-of-plane DMI  just below the QCP $(D_{z}/J\sim 0.08)$ \cite{zor} and thus  remains a QSL \cite{tia}. The kagom\'e calcium-chromium oxide Ca$_{10}$Cr$_7$O$_{28}$   has also been proposed as a QSL material  and there is no evidence  of a  strong DMI \cite{balz1}.
In the current study, we shall consider an alternative source of inducing magnetic order in frustrated KAFMs. The approach will be based on irradiation of laser light on the magnetic insulators.  This approach  has attracted considerable attention as a possible  mechanism to engineer synthetic topological  systems \cite{foot3,foot4,foot5,gru,fot,jot,fla,we2,we4,we5, gol,buk,eck,eck1,du,du1,du2,stru,tak}.

 In this letter, we formulate the theory of laser-irradiated KAFMs based on the Holstein-Primakoff magnon picture.  However, the basic idea can also be extended to charge-neutral bosonic spinons  in a similar fashion. The main results of this letter  are as follows.   First, we show that  circularly or linearly  polarized laser light is capable of lifting the ``zero energy mode''  in the KAFMs, thereby inducing  magnetic order. The associated magnetic excitation modes exhibit gaps at various points in the Brillouin zone. By inspection of measured spin waves in iron jarosites \cite{men4a}, we are able to conclude that the circularly polarized laser-induced symmetry breaking interactions possess distinctive features that are  similar to the effects of intrinsic in-plane and out-of-plane DMI on kagom\'e magnetic insulators \cite{men1}.  By studying the Floquet-Bloch thermal Hall effect close to thermal equilibrium, we further  establish that  the   magnetic excitation modes in circularly polarized light are topological, whereas those of linearly polarized light  are not. 
  These results suggest that laser-irradiation can be considered as one of the possible ways to induce  magnetic  order and nontrivial topological magnetic excitation modes  in KAFMs. Second, we apply the theory of laser-irradiation  to KAFMs with an  intrinsic out-of-plane DMI. The results show that radiation can also tune the out-of-plane DMI in KAFMs and induce a possible synthetic noncoplanar spin configuration with nonzero scalar spin chirality. 
  
\begin{figure}
\includegraphics[width=3in]{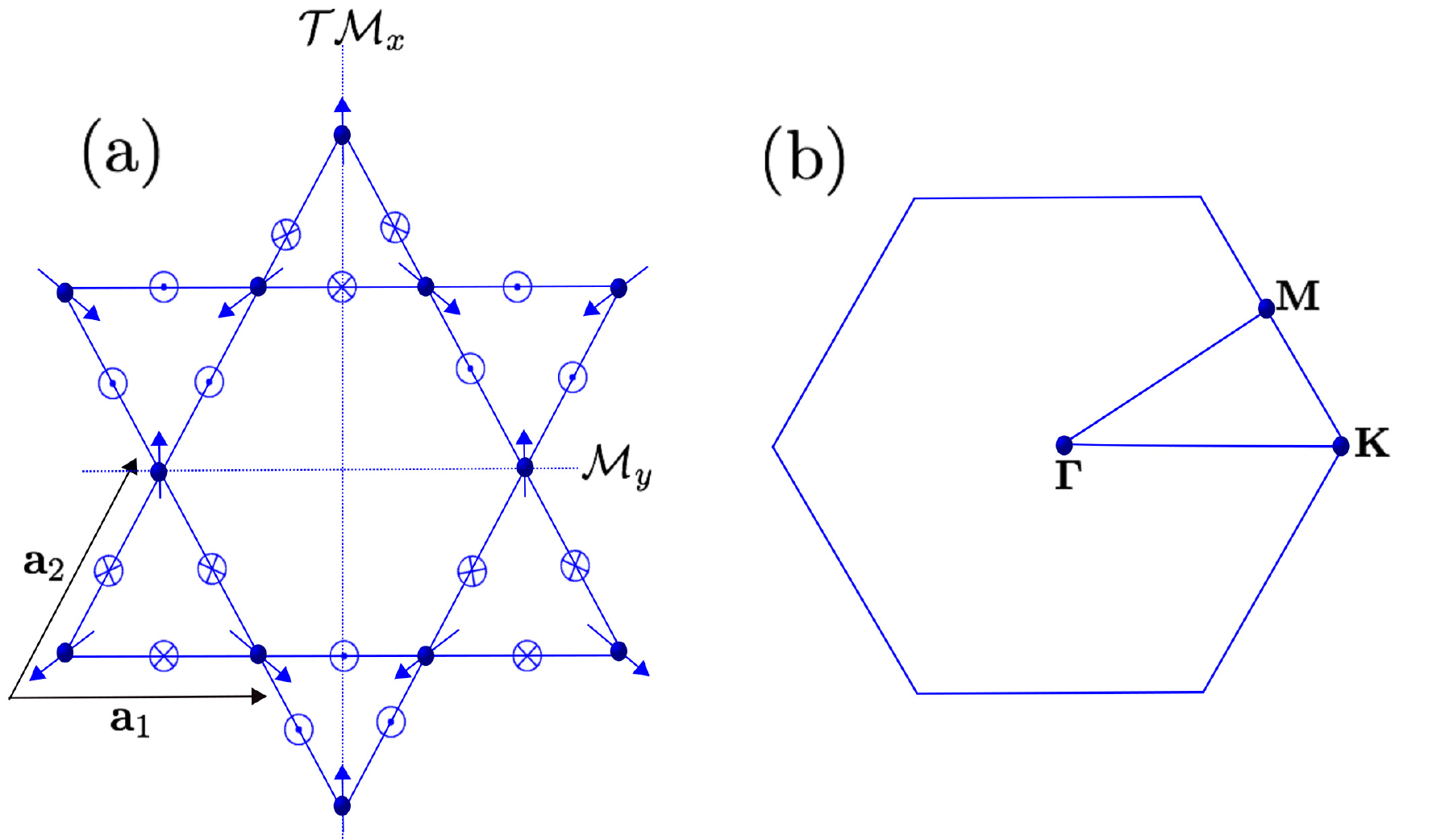}
\caption{ (a). Schematic of the kagom\'e lattice with out-of-plane DMI (dotted and crossed circles) and a coplanar ${\bf q=0}$ spin configuration (arrows). We also show the  mirror reflection symmetry $\mathcal{M}_y$ about the $y$-axis and $\mathcal T\mathcal M_x$ about the $x$-axis. The primitive vectors are ${\bf a}_1=(1,0)$, ${\bf a}_2=(1/2,\sqrt{3}/2)$, and ${\bf a}_3={\bf a}_2-{\bf a}_1$. (b) First Brillouin zone of kagom\'e lattice with indicated paths.}
\label{lat}
\end{figure}

Let us consider the simple Hamiltonian  for frustrated KAFMs, which is given by \begin{align}
\mathcal H&= J\sum_{\la ij\ra}  {\bf S}_{i}\cdot{\bf S}_{j}+ \mathcal H_{ani},
\label{h}
\end{align}
% \sum_{\la ij\ra} { \bf D}_{ij}\cdot{ \bf S}_{i}\times{\bf S}_{j}
where  ${\bf S}_i$ are the spin magnetic moments at the lattice sites $i$ located at ${\bf r}_i$ and $J>0$ is an antiferromagnetic interaction between nearest-neighbour (NN) sites. Here,  $ \mathcal H_{ani}$ is a small perturbative anisotropy to the Heisenberg exchange term, which is dominated by the out-of-plane DMI given by $ \mathcal H_{ani}=\sum_{\la ij\ra} { \bf D}_{ij}\cdot{ \bf S}_{i}\times{\bf S}_{j}$, where ${ \bf D}_{ij}=\pm D_z\hat z$ is an intrinsic out-of-plane DM component  due to inversion symmetry breaking  on the kagom\'e lattice at the midpoint connecting two magnetic sites, and the $\pm$ sign alternates between the triangular plaquettes of the kagom\'e lattice as shown in Fig.~\ref{lat}.

 In the absence of the out-of-plane DMI the classical ground states of the ideal Heisenberg KLAFM ({\it i.e.}, first term in Eq.\eqref{h})  are the ${\bf q =0}$ spin configurations shown in Fig.~\ref{lat}. However, they are infinitely degenerate. A direct application of linear spin wave approximation about this classical  spin configurations leads to a  ``zero energy mode'' \cite{har1}, which points to the fact that no particular ground-state spin configuration is favoured.  Application of a static Zeeman magnetic field partially lifts the degeneracy but does not remove it entirely. In the quantum limit this would imply that the system is disordered \cite{ran}. In the later sections, we will consider the possibility of lifting the ``zero energy mode'' and inducing magnetic order by laser-irradiation. 

 As previously shown the out-of-plane DMI induces and stabilizes the ${\bf q =0}$  classical spin  configurations \cite{men1}. The signs of the out-of-plane DMI determines which vector chirality of this long-range  magnetic  order is selected. In this report, we consider the  positive vector chirality $D_z>0$ with the minus sign. This form of the out-of-plane DMI respects the symmetries of the kagom\'e lattice in Fig.~\eqref{lat}. In particular, the combination of  time-reversal symmetry (TRS) $\mathcal T$ and mirror reflection symmetry $\mathcal M$ (i.e. $\mathcal T\mathcal M_x \mathcal T $ or $\mathcal M_y \mathcal T $ ), is a good symmetry of the coplanar ${\bf q =0}$ spin configuration. Therefore, we expect the underlying magnetic excitations to be protected by this symmetry and there should be a possibility of Dirac point in the Brillouin zone. As we will show later irradiation by laser light will modify this spin structure.

 The concept of laser-driven magnetic insulators  rely on the magnetic dipole moments of the underlying magnetic excitations as recently introduced in quantum ferromagnets \cite{owe}. This is due to the fact that  charge-neutral bosonic quasi-particles do not interact with an electromagnetic field except through their magnetic dipole moment. Therefore, this concept applies to both magnons and spinons.  We take the magnetic dipole moment to be along the in-plane ordering direction $\vec{ \mu}=-g\mu_B \hat{y}$, where $\mu_B$ is the Bohr magneton and $g$ is the spin $g$-factor. Now, we suppose that an in-plane laser light with dominant electric field components  $\bold{E}(t)$
is  irradiated on the kagom\'e lattice.  In the  background of the  time-varying electric field the hopping of charge-neutral bosonic quasi-particles will lead to  a time-dependent Aharonov-Casher phase \cite{aha}, which is given by
\begin{align}
\theta_{ij}(t)=\frac{g\mu_B}{\hbar }\int_{\bold{r}_i}^{\bold{r}_j} \bold A(t)\cdot d \boldsymbol{\ell},
\label{pha}
\end{align}
 where  ${\bf A}(t)=A_0(\sin\omega t, \sin(\omega t+\phi), 0)$ is the vector potential with amplitude $A_0$ due to the electric field  $\bold E(t)=-\frac{1}{c}\partial{\bold A(t)}/\partial t$. Note that circularly polarized laser light corresponds to $\phi=\pi/2$ and linearly polarization corresponds to $\phi=0$ or $\pi$.  In the following  we take the units  $\hbar=c=g\mu_B=1$.
 
 Next, we perform the standard spin wave analysis of the coplanar/noncollinear spin configuration  on the kagom\'e lattice \cite{har1,che}. The basic procedure follows by rotating the coordinate axis about the $z$-axis by the spin orientation angles: 
 
 \begin{align}
\mathcal{R}_z(\theta_i)
=\begin{pmatrix}
\cos\theta_i & -\sin\theta_i & 0\\
\sin\theta_i & \cos\theta_i &0\\
0 & 0 &1
\end{pmatrix},
\label{rot}
\end{align}
where $\theta_i=0,\pm 2\pi/3$. Hence the spin transforms as ${\bf S}_i= \mathcal{R}_z(\theta_i){\bf S}_i^\prime$. Next, we implement the  Holstein-Primakoff  transformation:  $S_{i}^{\prime y}\to S-a_{i}^\dagger a_{i},~ S_{i}^{\prime +}\to  \sqrt{2S}a_{i}=(S_{i}^{\prime-})^\dg$,  
 where $S^{\prime\pm} = S^{\prime x}\pm i S^{\prime z}$, and $a_{i}^\dagger(a_{i})$ are the bosonic creation (annihilation) operators. The corresponding time-dependent magnon tight-binding Hamiltonian is given by

\begin{align}
\mathcal H(t)&= JS\sum_{\la ij\ra}\Big[G_{ij}^0(a_i^\dg a_i +a_j^\dg a_j) +G_{ij}^1 (a_i^\dg a_j e^{i\theta_{ij}(t)} + h.c.)\label{eqnt} \\&\nonumber +G_{ij}^{2}(a_i^\dg a_j^\dg e^{i\theta_{ij}(t)} + h.c.)\Big],
\end{align}
where  $G_{ij}^0=(1+D_J)/2;~G_{ij}^1=(1-D_J)/4;~G_{ij}^2=(3+D_J)/4$, and $D_J=\sqrt{3}D_z/J$. Note that  the  hopping terms have now acquired a time-dependent phase by virtue of the Peierls substitution. In the momentum space we have $\mathcal H(t)=\sum_{\bo} \psi^\dg_\bo \mathcal{H}_\bo(t)\psi_\bo$, where $\psi^\dg_\bo= (a_{\bo 1}^{\dg},\thinspace a_{\bo 2}^{\dg},\thinspace a_{\bo 3}^{\dg}, \thinspace a_{-\bo 1},\thinspace a_{-\bo 2},\thinspace a_{-\bo 3} )$ is  the basis vector.

 \begin{align}
& \mathcal{H}_\bo(t)= 2JS\begin{pmatrix}
  {\boldsymbol{\mathcal{G}}^{0}}+\boldsymbol{\mathcal{G}}^1(t)& \boldsymbol{\mathcal{G}}^2(t)\\
\boldsymbol{\mathcal{G}}^2(t) &\  {\boldsymbol{\mathcal{G}}^{0}}+\boldsymbol{\mathcal{G}}^1(t)
\end{pmatrix},
\label{eqnr}
\end{align}
where ${\boldsymbol{\mathcal{G}}^{0}}=(1+D_J){\bf I}_{3\times 3}$, ${\boldsymbol{\mathcal{G}}^{1}}(t)=(1-D_J){\boldsymbol{\Lambda}}(t)/4$, and ${\boldsymbol{\mathcal{G}}^{2}}(t)=(3+D_J){\boldsymbol{\Lambda}}(t)/4$. \begin{widetext}
\begin{align}
&\boldsymbol{\Lambda}(t)=\begin{pmatrix}
  0& \cos[\lb\bo + {\bf A}(t)\rb\cdot {\bf a}_1] &\cos[\lb\bo + {\bf A}(t)\rb\cdot {\bf a}_2] \\
\cos[\lb\bo + {\bf A}(t)\rb\cdot {\bf a}_1] & 0&\cos[\lb\bo + {\bf A}(t)\rb\cdot {\bf a}_3]\\
\cos[\lb\bo + {\bf A}(t)\rb\cdot {\bf a}_2] & \cos[\lb\bo + {\bf A}(t)\rb\cdot {\bf a}_3] & 0 \\  
 \end{pmatrix},
\end{align}
\end{widetext}
 Note that Eqs.~\eqref{eqnt} and \eqref{eqnr} have off-diagonal terms which do not exist in ferromagnets \cite{owe}. However, the general formalism of Floquet theory applies to any time-dependent  tight-binding Hamiltonian.

\begin{figure}
\centering
\includegraphics[width=1\linewidth]{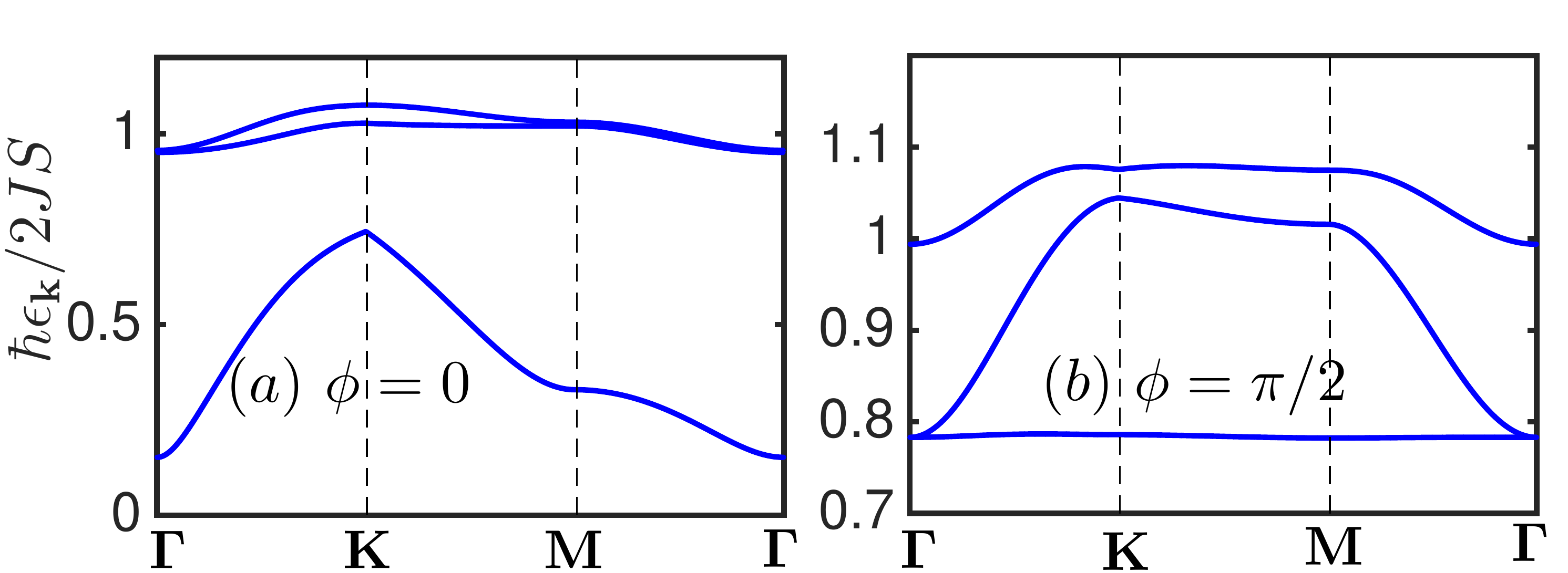}
\includegraphics[width=1\linewidth]{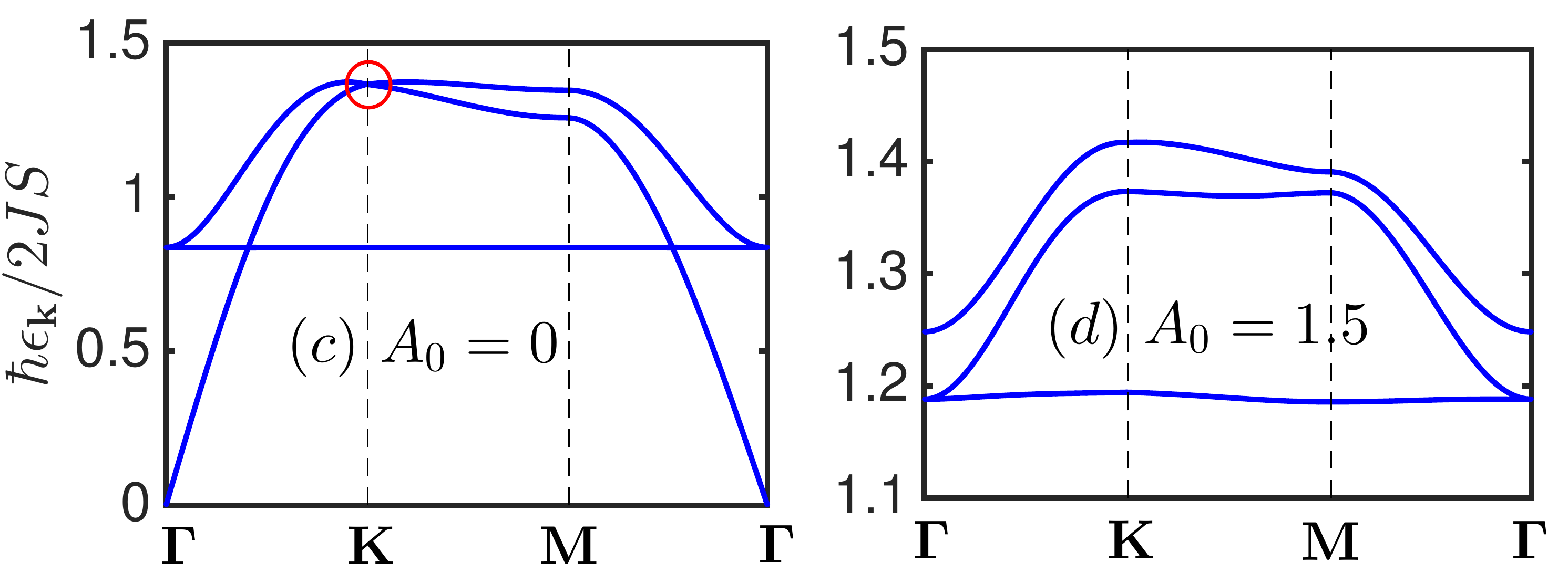}
\caption{Color online. Top figure. Floquet-Bloch magnon bands of frustrated  KAFM for  $~D_z/J=0$. (a) Linearly polarized light. (b) Circularly polarized light. Here, $\omega/J=10, A_0=1.5$. Bottom figure. Floquet-Bloch magnon bands of frustrated  KAFM for $~D_z/J=0.2$. (c) Undriven system.  The circled point is a Dirac magnon node. (d) Laser-driven system by circularly polarized light $\phi=\pi/2$, with $\omega/J=10$.}
\label{band}
\end{figure}

 The basic idea of Floquet theory is to  transform a time-dependent Hamiltonian such as Eq.~\eqref{eqnt} or \eqref{eqnr}  into an effective static Hamiltonian.  To proceed, we write the Floquet-Bloch wave function that obey time-dependent Schr\"{o}dinger equation  as $\Psi_{\bo\alpha}(t)=e^{i\epsilon_{\bo\alpha}(t)}\Phi_{\bo\alpha}(t)$,
where $\Phi_{\bo\alpha}(t)=\Phi_{\bo\alpha}(t +T)$ is a periodic function that denotes the Floquet-Bloch states for band $\alpha$ with  period $T=2\pi/\omega$, and $\epsilon_{\bo\alpha}(t)$ is the quasi-energy. The periodic function can be expanded in Fourier space: $
\Phi_{\bo\alpha}(t)=\sum_{m}e^{im\omega t}\Phi_{\bo\alpha}^m$. We define the Floquet Hamiltonian operator as $\mathcal H_{\bo}^F(t)=\mathcal H_{\bo}(t)-i\partial/\partial t$. Then the states $\Phi_{\bo\alpha}^m$ leads to a time-independent Floquet energy eigenvalue equation
\begin{align}
\sum_m [\mathcal H_\bo^{n-m} + m\omega\delta_{n,m}]\Phi_{\bo\alpha}^m=\epsilon_{\bo\alpha}\Phi_{\bo\alpha}^n,
\end{align}
where $\mathcal H_\bo^{\ell }=\frac{1}{T}\int_0^T dt e^{-i\ell\omega t}\mathcal H_\bo(t)$. 

The Floquet formalism is reliable in the high frequency limit $\omega\gg J$. Therefore, we work in this limit  and consider the truncation $\ell=0,\pm 1$. At this juncture the Floquet Hamiltonian contains the zeroth order Bessel function $\mathcal J_0(\xi)$ and  the first order Bessel function $\mathcal J_1(\xi)$, where $\xi$ depends on the amplitude $A_0$ and phase $\phi$. However, the resulting Hamiltonian is a big matrix comprising numerous Floquet-Bloch side-bands. Therefore, analytical analysis is unfeasible we therefore resort to numerical analysis. We first consider the limit of zero out-of-plane DMI ($D_z/J=0$). In this case a ``zero energy mode'' is inevitable \cite{har1}. In Fig.~\eqref{band} we have shown the Floquet-Bloch magnon bands  in the presence of linearly (a) and circularly (b) polarized light  for $D_z/J=0$ and $A_0=1.5$ along the Brillouin zone paths \cite{owee} in Fig.~\eqref{lat}. In both cases, it is evident that the ``zero energy mode'' is lifted to nearly flat mode at nonzero energy. We also observe that all the bands have a finite energy gap at the ${\bf \Gamma}$-point as well as the ${\bf K}$-point. 
 
It is believed  that circularly polarized laser light breaks TRS, but it is very crucial to identify the form of the  symmetry breaking interaction induced by the laser light. Interestingly,   the magnon bands in Fig.~\ref{band} (b) for circularly  polarized light exhibit all the important features reported in  kagom\'e iron jarosites \cite{men4a}. They include  gapped magnetic excitations at various points in the Brillouin zone, which were theoretically identified as a consequence of the out-of-plane and in-plane intrinsic DMIs \cite{men4a}. Therefore, we infer that in the driven KAFMs circularly polarized light induces synthetic out-of-plane and in-plane DMIs. The former  is responsible for the stability of the coplanar ${\bf q=0}$ spin configuration leading to lifted ``zero energy mode'' as shown in Fig.~\ref{lat}. Whereas the latter which points inside the triangular plaquettes \cite{sup1a} breaks mirror reflection symmetry and rotational invariance and it is responsible for spin canting out-of-plane and leads to  a non-coplanar umbrella spin configuration with nonzero scalar spin chirality.   The in-plane DMI is the primary source  of the energy gaps at the ${\bf \Gamma}$ and  ${\bf K}$ points. Therefore we conclude that irradiation by laser light can induce long-range magnetic  order in highly frustrated magnets in the same way as the intrinsic DMIs \cite{men1}.  As we will discuss in the following the laser-induced anisotropy in the case of  linearly polarized light [Fig.~\ref{band} (a)]  does not break TRS.

Next, we consider the limit of nonzero out-of-plane DMI ($D_z/J\neq 0$). As shown in Fig.~\ref{band} (c) the intrinsic out-of-plane DMI lifts the ``zero energy mode'' in the undriven system and stabilizes the conventional  coplanar ${\bf q=0}$ spin configuration with a Dirac magnon node at the ${\bf K}$-point protected by  the effective TRS $\mathcal T\mathcal M_y$. We can also see a gapless Goldstone mode at the ${\bf \Gamma}$-point due to U(1) rotational invariance.  As the laser drive is tuned on this effective TRS is bound to be broken and the Dirac magnon node will be lifted as shown in Fig.~\ref{band} (d). The presence of energy gap at the ${\bf \Gamma}$-point also suggests that the induced synthetic anisotropy is an in-plane DM component which obviously breaks rotational invariance and modifies the conventional  ${\bf q=0}$ spin configuration as previously mentioned. 

 Now, we would like to compute an experimentally observable quantity in DM models on the kagom\'e lattice. The gapped excitations at the ${\bf K}$-point suggests that the Floquet-Bloch  magnon bands will be topologically nontrivial with finite Berry curvature. The effects of topologically nontrivial magnetic excitations are believed to manifest in the study of thermal Hall effect.  Recently, the study of thermal Hall effect of charge-neutral bosonic quasi-particles has attracted a lot of interest in ferromagnets \cite{alex1, alex1a, alex6,s1,s2,s5, alex4, alex5a, shi, sol1} and spin liquid magnetic insulators \cite{wat, hir}. In the former, the theory of Berry curvature induced by the DMI  can explain the observed  thermal Hall conductivity. In the latter, however, the origin of thermal Hall conductivity remains an open question, but the  scalar spin chirality could play a vital role \cite{sol2}. 

\begin{figure}
\centering
\includegraphics[width=1.1\linewidth]{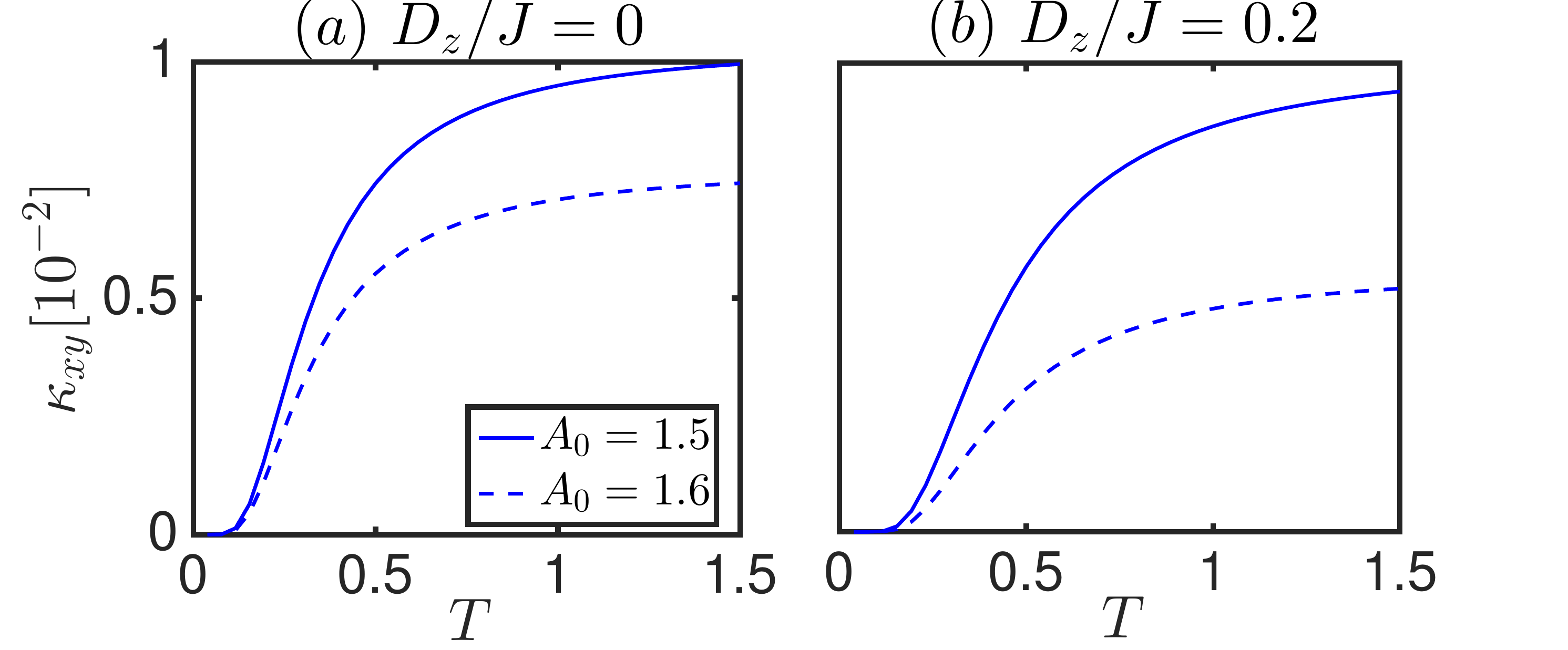}
\caption{Color online.  Floquet thermal Hall conductivity vs. temperature for circularly polarized light $\phi=\pi/2$ and two laser amplitudes with $\omega/J=10$.}
\label{THE}
\end{figure}

In the current study,   synthetic  TRS breaking interactions are induced by circularly polarized laser light and they are identified as the in-plane and out-of-plane DMI. Thus, the theory of thermal Hall effect can be applied in a similar way. However, the Floquet theory leads to nonequilibrium distribution of the quasi-particles  and the Bose function $n_B(\epsilon_{\bo\alpha})$ will depend on the detail properties of the system. In the following we focus on the case in which the Bose function is close to thermal equilibrium and write the  linear response thermal Hall conductivity as \cite{s5} $\kappa_{xy}=-k_B^2 T\int_{{BZ}} \frac{d^2k}{(2\pi)^2}~ \sum_{\alpha=1}^N c_2\lb n_\alpha\rb\Omega_{\alpha\bo},$
where $n_\alpha\equiv n_B(\epsilon_{\alpha\bo})=(e^{{\epsilon_{\alpha\bo}}/k_BT}-1)^{-1}$ is the Bose function close to thermal equilibrium, $c_2(x)=(1+x)\lb \ln \frac{1+x}{x}\rb^2-(\ln x)^2-2\textrm{Li}_2(-x),$ and $\text{Li}_n(x)$ is a polylogarithm.  The Berry curvature is given by $\Omega_{\alpha \bo}=\lb\boldsymbol{\nabla}\times \boldsymbol{\mathcal A}_{\alpha\bo}\rb_z$,  where $\boldsymbol{\mathcal A}_{\alpha\bo}=i\braket{\Phi_{\alpha\bo}|\boldsymbol{\nabla}|\Phi_{\alpha\bo}}$ is the Berry connection. 

 We note that $\Omega_{\alpha\bo}$ vanishes in the undriven system with only out-of-plane DMI due to an effective TRS, and hence $\kappa_{xy}$ is zero. For laser-driven system by circularly polarized light  the Chern number defined as the integration of the Berry curvature over the Brillouin zone is proportional to the synthetic scalar-chirality of the noncoplanar spin configurations induced by the synthetic in-plane DMI. A nonzero thermal Hall conductivity implies that the underlying magnetic excitations are topologically nontrivial. In Fig.~\ref{THE} we have confirmed that, indeed, the magnetic excitations are topologically nontrivial in the case of circularly polarized light ($\phi=\pi/2$) with nonzero $\kappa_{xy}$ for $D_z/J=0$ (a) and $D_z/J=0.2$ (b). Furthermore, we have checked  numerically that $\kappa_{xy}$ vanishes for linearly polarized light ($\phi=0$ or $\pi$) --- an indication that the laser-induced anisotropy in this case does not break TRS.

% It was recently introduced to quantum ferromagnets, where a  synthetic out-of-plane DMI is explicitly induced by the electric field of the laser light  \cite{owe}. In frustrated quantum magnets, however, the out-of-plane DMI does not lead to topological magnetic excitations unlike in quantum ferromagnets \cite{alex1, alex1a, alex6,xc,sol,s1,s2,s4,s5, alex4, alex5a, lifa, shi, sol1, mok, su, su1}. Instead, it induces a  long-range magnetic ${\bf q =0}$  order \cite{men1}, which preserves an effective TRS with vanishing Berry curvature. In fact, the mechanism that leads to topological magnetic excitations with finite thermal Hall effect in spin liquid magnetic insulators is still an open question \cite{wat, hir}. 

In summary, we have shown that the dispersionless   ``zero energy mode'' in  frustrated KAFMs can be lifted by laser-irradiation and long-range magnetic order can be induced.  The possible circularly polarized laser-induced synthetic interactions are identified as the in-plane and out-of-plane DMIs, which play the same role in frustrated KAFMs with low crystal symmetry \cite{men1}. In the present case, however,  their strength  can be tuned by the laser light.  We also showed that the system possessed transport properties that can be experimentally accessible. We believe  that these results are within experimental reach and can be accessible with the current terahertz frequency using ultrafast terahertz spectroscopy \cite{pri}. Thus far, induced magnetic order in frustrated magnets is limited to applying an external magnetic field or pressure. However, these methods usually do not lead to an induced in-plane and out-of-plane DMI.  Therefore, experiments can now look for how laser irradiation modifies the properties of frustrated magnets. Most importantly, the laser-induced topological magnetic excitation should be the primary motive as it quantifies the effects of laser irradiation on the system. Then thermal Hall effect can be measured by applying a temperature gradient.   Currently, a lifted ``zero energy mode'' has only been seen clearly in iron jarosite \cite{men4a}, but there are numerous frustrated kagom\'e antiferromagnets.   Therefore, another experimental task would be the possibility of  inducing lifted ``zero energy mode'' by  laser irradiation on frustrated kagom\'e antiferromagnets such as herbertsmithite ZnCu$_3$(OH)$_6$Cl$_2$, which is a known  kagom\'e antiferromagnet with QSL properties. In the future, it would be interesting to extend the current approach to spinons by the Schwinger boson formalism. This will pave the way for optical manipulation of strongly correlated materials including chiral spin liquids \cite{mat}.

 Research at Perimeter Institute is supported by the Government of Canada through Industry Canada and by the Province of Ontario through the Ministry of Research and Innovation.

\end{document}